# Serializable Snapshot Isolation in PostgreSQL

Dan R. K. Ports
MIT CSAIL
32 Vassar St.
Cambridge, MA 02139
drkp@csail.mit.edu

Kevin Grittner
Consolidated Court Automation Programs
Wisconsin Supreme Court
110 East Main Street
Madison, WI 53703
Kevin.Grittner@wicourts.gov

## ABSTRACT

This paper describes our experience implementing PostgreSQL's new serializable isolation level. It is based on the recently-developed Serializable Snapshot Isolation (SSI) technique. This is the first implementation of SSI in a production database release as well as the first in a database that did not previously have a lock-based serializable isolation level. We reflect on our experience and describe how we overcame some of the resulting challenges, including the implementation of a new lock manager, a technique for ensuring memory usage is bounded, and integration with other PostgreSQL features. We also introduce an extension to SSI that improves performance for read-only transactions. We evaluate PostgreSQL's serializable isolation level using several benchmarks and show that it achieves performance only slightly below that of snapshot isolation, and significantly outperforms the traditional two-phase locking approach on read-intensive workloads.

## 1. OVERVIEW

Serializable isolation for transactions is an important property: it allows application developers to write transactions as though they will execute sequentially, without regard for interactions with concurrently-executing transactions. Until recently, PostgreSQL, a popular open-source database, did not provide a serializable isolation level because the standard two-phase locking mechanism was seen as too expensive. Its highest isolation level was snapshot isolation, which offers greater performance but allows certain anomalies.

In the latest PostgreSQL 9.1 release,[1] we introduced a serializable isolation level that retains many of the performance benefits of snapshot isolation while still guaranteeing true serializability. It uses an extension of the Serializable Snapshot Isolation (SSI) technique from current research [7]. SSI runs transactions using snapshot isolation, but checks at runtime for conflicts between concurrent transactions, and aborts transactions when anomalies are possible. We extended SSI to improve performance for read-only transactions, an important part of many workloads.

---

[1] PostgreSQL 9.1 is available for download from http://www.postgresql.org/.

This paper describes our experiences implementing SSI in PostgreSQL. Our experience is noteworthy for several reasons:

It is the first implementation of SSI in a production database release. Accordingly, it must address interactions with other database features that previous research prototypes have ignored. For example, we had to integrate SSI with PostgreSQL's support for replication systems, two-phase commit, and subtransactions. We also address memory usage limitations, an important practical concern; we describe a *transaction summarization* technique that ensures that the SSI implementation uses a bounded amount of RAM without limiting the number of concurrent transactions.

Ours is also the first implementation of SSI for a purely snapshot-based DBMS. Although SSI seems especially suited for such databases, earlier SSI implementations were based on databases that already supported serializable isolation via two-phase locking, such as MySQL. As a result, they were able to take advantage of existing predicate locking mechanisms to detect conflicting transactions for SSI. Lacking this infrastructure, we were required to build a new lock manager. Our lock manager is specifically optimized for tracking SSI read dependencies, making it simpler in some respects than a classic lock manager but also introducing some unusual challenges. PostgreSQL 9.1 uses this lock manager, along with multiversion concurrency control data, to detect conflicts between concurrent transactions. We also introduce a *safe retry* rule, which resolves conflicts by aborting transactions in such a way that an immediately retried transaction does not fail in the same way.

Read-only transactions are common, so PostgreSQL 9.1 optimizes for them. We extend SSI by deriving a result in multiversion serializability theory and applying it to reduce the rate of false positive serialization failures. We also introduce *safe snapshots* and *deferrable transactions*, which allow certain read-only transactions to execute without the overhead of SSI by identifying cases where snapshot isolation anomalies cannot occur.

PostgreSQL 9.1's serializable isolation level is effective: it provides true serializability but allows more concurrency than two-phase locking. Our experiments with a transaction processing and a web application benchmark show that our serializable mode has a performance cost of less than 7% relative to snapshot isolation, and outperforms two-phase locking significantly on some workloads.

This paper begins with an explanation of how snapshot isolation differs from serializability and why we view serializability as an important DBMS feature in Section 2. Section 3 describes the SSI technique and reviews the previous work. Section 4 extends SSI with new optimizations for read-only transactions. We then turn to the implementation of SSI in PostgreSQL 9.1, with Section 5 giving an overview of the implementation and Section 6 discussing techniques for reducing its memory usage. Section 7 examines how SSI interacts with other PostgreSQL features. Finally, in Section 8 we compare the performance of our implementation to PostgreSQL's existing snapshot isolation level and to a lock-based implementation of serializability.

## 2. SNAPSHOT ISOLATION VERSUS SERIALIZABILITY

Users group operations into transactions to ensure they are atomic with respect to other concurrently-executing transactions, as well as with respect to crashes. ANSI SQL allows the user to request one of several isolation levels; the strongest is serializability [4]. In a serializable execution, the effects of transactions must be equivalent to executing the transactions in some serial order. This property is appealing for users because it means that transactions can be treated in isolation: if each transaction can be shown to do the right thing when run alone (such as maintaining a data integrity invariant), then it also does so in *any* mix of concurrent transactions.

At weaker isolation levels, race conditions between concurrent transactions can produce a result that does not correspond to any serializable execution. In spite of this, such isolation levels can provide better performance and are commonly used. For example, PostgreSQL, like many other databases, uses its weakest isolation level, READ COMMITTED, by default. This level guarantees only that transactions do not see uncommitted data, but offers high performance because it can be implemented without read locks in PostgreSQL's multiversion storage system; a lock-based DBMS could implement it using only short-duration read locks [12].

Snapshot isolation (SI) is one particular weak isolation level that can be implemented efficiently using multiversion concurrency control, without read locking. It was previously the strongest isolation level available in PostgreSQL: users requesting SERIALIZABLE mode actually received snapshot isolation (as they still do in the Oracle DBMS). However, snapshot isolation does not guarantee serializable behavior; it allows certain anomalies [2,5]. This unexpected transaction behavior can pose a problem for users that demand data integrity, such as the Wisconsin Court System, one of the motivating cases for this work. In particular, snapshot isolation anomalies are difficult to deal with because they typically manifest themselves as silent data corruption (*e.g.* lost updates). In many cases, the invalid data is not discovered until much later, and the error cannot easily be reproduced, making the cause difficult to track down.

### 2.1 Snapshot Isolation

In snapshot isolation, all reads within a transaction see a consistent view of the database, as though the transaction operates on a private snapshot of the database taken before its first read. Section 5.1 describes how PostgreSQL implements these snapshots using versioned tuples. In addition, SI prohibits concurrent transactions from modifying the same data. Like most SI databases, PostgreSQL uses tuple-level write locks to implement this restriction.

Snapshot isolation does not allow the three anomalies defined in the ANSI SQL standard: dirty reads, non-repeatable reads, and phantom reads. However, it allows several other anomalies. These anomalies were not initially well understood, and they remain poorly understood in practice. For example, there is a common misconception that avoiding the aforementioned three anomalies is a sufficient condition for serializability, and for years the PostgreSQL documentation did not acknowledge the difference between its SERIALIZABLE mode and true serializability.

#### 2.1.1 Example 1: Simple Write Skew

The simplest anomaly occurs between two concurrent transactions that read the same data, but modify disjoint sets of data. Consider the two transactions in Figure 1 (based on an example given by Cahill et al [7]). Each checks whether there are at least two doctors on call, and if so takes one doctor off call. Given an initial state where Alice and Bob are the only doctors on call, it can easily be verified that executing $T_1$ and $T_2$ sequentially in either order will leave at least one doctor on call – making these transactions an effective way of enforcing that invariant.

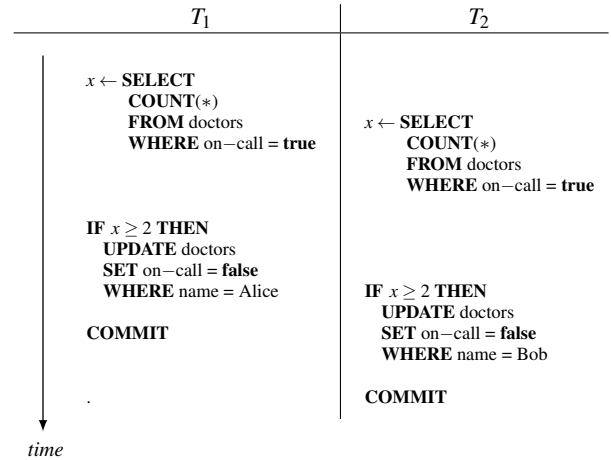

Figure 1: A simple write-skew anomaly

But the interleaving of Figure 1, when executed under snapshot isolation, violates that invariant. Both transactions read from a snapshot taken when they start, showing both doctors on call. Seeing that, they both proceed to remove Alice and Bob, respectively, from call status. The write locks taken on update don't solve this problem, because the two transactions modify different rows and thus do not conflict. By contrast, in two-phase locking DBMS, each transaction would take read locks that would conflict with the other's write. Similarly, in an optimistic-concurrency system, the second transaction would fail to commit because its read set is no longer up to date.

#### 2.1.2 Example 2: Batch Processing

The previous example is simple in the sense that it consists of only two transactions that directly conflict with each other. But more complex interactions between transactions are possible. Here, we give an example of a snapshot isolation anomaly resulting from three transactions, one of which is read-only.

Consider a transaction-processing system that maintains two tables. A *receipts* table tracks the day's receipts, with each row tagged with the associated batch number. A separate *control* table simply holds the current batch number. There are three transaction types:

- NEW-RECEIPT: reads the current batch number from the control table, then inserts a new entry in the receipts table tagged with that batch number
- CLOSE-BATCH: increments the current batch number in the control table
- REPORT: reads the current batch number from the control table, then reads all entries from the receipts table with the *previous* batch number (*i.e.* to display a total of the previous day's receipts)

The following useful invariant holds under serializable executions: after a REPORT transaction has shown the total for a particular batch, subsequent transactions cannot change that total. This is because the REPORT shows the previous batch's transactions, so it must follow a CLOSE-BATCH transaction. Every NEW-RECEIPT transaction must either precede both transactions, making it visible to the REPORT,

| $T_1$ (REPORT) | $T_2$ (NEW-RECEIPT) | $T_3$ (CLOSE-BATCH) |
|---|---|---|
| | $x \leftarrow$ **SELECT** current_batch | |
| | | **INCREMENT** current_batch |
| | | **COMMIT** |
| $x \leftarrow$ **SELECT** current_batch | | |
| **SELECT SUM**(amount) **FROM** receipts **WHERE** batch = $x - 1$ | | |
| **COMMIT** | | |
| | **INSERT INTO** receipts **VALUES** (x, somedata) | |
| . | **COMMIT** | . |

Figure 2: An anomaly involving three transactions

or follow the CLOSE-BATCH transaction, in which case it will be assigned the next batch number.

However, the interleaving shown in Figure 2 is allowed under SI, and violates this invariant. The receipt inserted by transaction $T_2$ has the previous batch number because $T_2$ starts before $T_3$ commits, but is not visible in the corresponding report produced by $T_1$.

Interestingly, this anomaly requires all three transactions, including $T_1$ – even though it is read-only. Without it, the execution *is* serializable, with the serial ordering being $\langle T_2, T_3 \rangle$. The fact that read-only transactions can be involved in SI anomalies was a surprising result discovered by Fekete et al. [11].

## 2.2 Why Serializability?

Snapshot isolation anomalies like those described above are undesirable because they can cause unexpected transaction behavior that leads to inconsistencies in the database. Nevertheless, SI is widely used, and many techniques have been developed to avoid anomalies:

- some workloads simply don't experience any anomalies; their behavior is serializable under snapshot isolation. The TPC-C benchmark is one such example [10].

- if a potential conflict between two transactions is identified, explicit locking can be used to avoid it. PostgreSQL provides explicit locking at the table level via the LOCK TABLE command, and at the tuple level via SELECT FOR UPDATE.

- alternatively, the conflict can be *materialized* by creating a dummy row to represent the conflict, and forcing every transaction involved to update that row [10].

- if the desired goal is to enforce an integrity constraint, and that constraint can be expressed to the DBMS (*e.g.* using a foreign key, uniqueness, or exclusion constraint), then the DBMS can enforce it regardless of isolation level.

Given the existence of these techniques, one might question the need to provide serializable isolation in the database: shouldn't users just program their applications to handle the lower isolation level? (We have often been asked this question.) Our view is that providing serializability in the database is an important simplification for application developers, because concurrency issues are notoriously difficult to deal with. Indeed, SI anomalies have been discovered in real-world applications [14]. The analysis required to identify potential anomalies (or prove that none exist) is complex and is likely beyond the reach of many users. In contrast, serializable transactions offer simple semantics: users can treat their transactions as though they were running in isolation.

In particular, the analysis is difficult because it inherently concerns interactions *between* transactions. Thus, each transaction must be analyzed in the context of all other transactions that it might run concurrently with. It is difficult to do this $n^2$ analysis in a dynamic environment with many complex transactions. Such was the case at the Wisconsin Court System. Data integrity is a critical concern, given the nature of the data (*e.g.* warrant status information) and regulatory requirements. Snapshot isolation anomalies posed a dangerous threat to data integrity, especially because they can cause silent corruption. At the same time, with a complex schema (hundreds of relations), over 20 full-time programmers writing new queries, and queries being auto-generated by object-relational frameworks, analyzing query interactions to find possible anomalies – and keeping the analysis up to date – was simply not feasible.

A further problem is that using static analysis to identify anomalies may not be possible when the workload includes ad hoc queries. Even applications that execute pre-defined stored procedures are likely to also have occasional ad hoc queries for administrative tasks. For example, an administrator might manually execute queries (*e.g.* using the psql command line utility or a front-end like pgAdmin) to inspect the database or repair corrupted data. Static analysis, lacking knowledge of these transactions, cannot prevent anomalies involving them. Even read-only ad hoc transactions, such as making a copy of the database with the pg_dump utility, can expose anomalous states of the database.

## 3. SERIALIZABLE SNAPSHOT ISOLATION

Our implementation of serializability in PostgreSQL is unique among production databases in that it uses the recently-developed Serializable Snapshot Isolation (SSI) technique [7]. Nearly all other databases that provide serializability do so using strict two-phase locking (S2PL). In S2PL, transactions acquire locks on all objects they read or write, and hold those locks until the transaction commits. To prevent phantoms, these locks must be predicate locks, usually implemented using index-range locks.

One could certainly implement a serializable isolation level for PostgreSQL using S2PL, but we did not want to do so for performance reasons. Indeed, the original POSTGRES storage manager inherited from the Berkeley research project had precisely that, using a conventional lock manager to provide concurrency control [19]; its replacement with a multiversion concurrency control (MVCC) system in 1999 was one of the first major accomplishments of the PostgreSQL open-source community. Subsequently, the benefits of MVCC have become firmly ingrained in the PostgreSQL ethos, making a lock-based SERIALIZABLE mode that behaved so differently a non-starter. Users accustomed to the "readers don't block writers, and writers don't block readers" mantra would be surprised by the additional blocking, and a S2PL approach was unpalatable to most of the developer community.

SSI takes a different approach to ensuring serializability: it runs transactions using snapshot isolation, but adds additional checks to determine whether anomalies are possible. This is based on a theory of snapshot isolation anomalies, discussed below. SSI was appealing to us because it built on snapshot isolation, and offered higher performance than a S2PL implementation. Another important factor was that SSI does not require any additional blocking. Transactions that might violate serializability are simply aborted. Because basic snapshot isolation can already roll back transactions due to update

1852

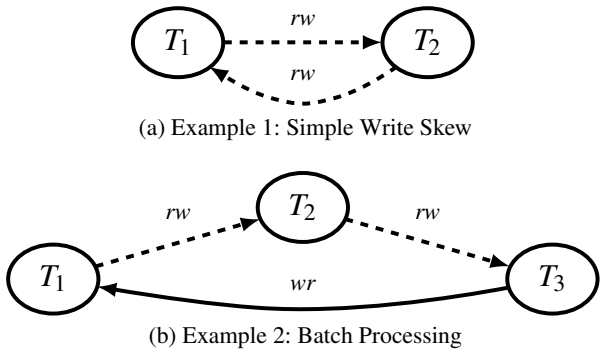

Figure 3: Serialization graphs for Examples 1 and 2

conflicts, users must already be prepared to handle transactions aborted by serialization failures, *e.g.* using a middleware layer that automatically retries transactions.

The remainder of this section reviews the previous work on SSI. Sections 3.1 and 3.2 review the theory of snapshot isolation anomalies and when they arise. Section 3.3 describes the SSI algorithm and some proposed variants on it.

## 3.1 Snapshot Isolation Anomalies

SSI builds on a line of research studying the nature of snapshot isolation anomalies. Not long after snapshot isolation was implemented, it was observed that SI allowed non-serializable executions but did not exhibit any of the well-understood anomalies proscribed by the SQL standard [5], This suggested that existing isolation level definitions were inadequate, and prompted an effort to define the anomalies caused by SI and when they arise.

Adya et al. [2] proposed representing an execution with a multiversion serialization history graph. This graph contains a node per transaction, and an edge from transaction $T_1$ to transaction $T_2$ if $T_1$ must have preceded $T_2$ in the apparent serial order of execution. Three types of dependencies can create these edges:

- **wr-dependencies**: if $T_1$ writes a version of an object, and $T_2$ reads that version, then $T_1$ appears to have executed before $T_2$
- **ww-dependencies**: if $T_1$ writes a version of some object, and $T_2$ replaces that version with the next version, then $T_1$ appears to have executed before $T_2$
- **rw-antidependencies**: if $T_1$ writes a version of some object, and $T_2$ reads the previous version of that object, then $T_1$ appears to have executed *after* $T_2$, because $T_2$ did not see its update. As we will see, these dependencies are central to SSI; we sometimes also refer to them as **rw-conflicts**.

If a cycle is present in the graph, then the execution does not correspond to any serial order, *i.e.* a snapshot isolation anomaly has caused a serializability violation. Otherwise, the serial order can be determined using a topological sort.

Note that the definitions above referred to *objects*. We use this more abstract term rather than "tuple" or "row" because dependencies can also be caused by predicate reads. For example, if $T_1$ scans a table for all rows where $x = 1$, and $T_2$ subsequently inserts a new row matching that predicate, then there is a $T_1 \xrightarrow{rw} T_2$ rw-antidependency.

Figure 3 shows the serialization graphs corresponding to Examples 1 and 2 from Section 2.1. For Example 1, $T_1$ updates the row containing Alice's call status, but this update is not visible to $T_2$'s SELECT, creating a rw-antidependency: $T_2$ appears to have executed before $T_1$. Similarly, $T_2$'s UPDATE is not visible to $T_1$, creating a rw-antidependency in the opposite direction. The resulting cycle indicates that the execution is not serializable. Example 2, the batch-processing example, contains three transactions and two types of dependencies. $T_3$, which increments the batch number, appears to execute after $T_2$, which reads the old version. The receipt inserted by $T_2$ does not appear in $T_1$'s report, so $T_2$ appears to execute after $T_1$. Finally, $T_3$ appears to execute *before* $T_1$, completing the cycle. This last edge is a wr-dependency: $T_3$'s increment of the batch number was visible to $T_1$'s read, because $T_3$ committed before $T_1$ began.

## 3.2 Serializability Theory

Note that a wr-dependency from *A* to *B* means that *A* must have committed before *B* began, as this is required for *A*'s changes to be visible to *B*'s snapshot. The same is true of ww-dependencies because of write locking. However, rw-antidependencies occur between *concurrent* transactions: one must start while the other was active. Therefore, they play an important role in SI anomalies.

Adya [1] observed that every cycle in the serialization graph (*i.e.* every anomaly) contains at least two rw-antidependency edges. Fekete et al. [10] subsequently showed that two such edges must be adjacent:

**Theorem 1** (Fekete et al. [10]). *Every cycle in the serialization history graph contains a sequence of edges $T_1 \xrightarrow{rw} T_2 \xrightarrow{rw} T_3$ where each edge is a rw-antidependency. Furthermore, $T_3$ must be the first transaction in the cycle to commit.*

Note that this is actually a stronger statement than that given by Fekete et al., who state only that $T_3$ must commit before $T_1$ and $T_2$. Though not explicitly stated, it is a consequence of their proof that $T_3$ must be the first transaction in the *entire cycle* to commit.

**Corollary 2.** *Transaction $T_1$ is concurrent with $T_2$, and $T_2$ is concurrent with $T_3$, because rw-antidependencies occur only between concurrent transactions.*

Note that $T_1$ and $T_3$ may refer to the same transaction, for cycles of length 2 such as the one in the write-skew example (Figure 3a).

## 3.3 SSI

Cahill et al. introduced SSI, a technique for providing serializability using snapshot isolation, by detecting potential anomalies at runtime, and aborting transactions as necessary [7]. It is similar to concurrency control protocols based on serialization graph testing [8], in that it tracks edges in the serialization graph and prevents cycles from forming. However, rather than testing the graph for cycles, it checks for a "dangerous structure" of two adjacent rw-antidependency edges. If any transaction has both an incoming rw-antidependency and an outgoing one, SSI aborts one of the transactions involved. Theorem 1 shows that doing so ensures serializable execution, but it may have false positives because not every dangerous structure is part of a cycle. The benefit is that it is more efficient. Besides being a less expensive runtime check than cycle testing, dangerous structures are composed entirely of rw-antidependencies, so SSI does not need to track wr- and ww-dependency edges.

This approach can offer greater concurrency than a typical S2PL or optimistic concurrency control (OCC) [15] system. Essentially, both S2PL and classic OCC prevent concurrent transactions from having rw-conflicts. SSI allows some rw-conflicts as long as they do not form a dangerous structure, a less restrictive requirement. For instance, consider Example 2 with the read-only transaction $T_1$ removed. We saw in Section 2.1.2 that this execution is serializable even though there is a rw-antidependency $T_2 \xrightarrow{rw} T_3$. However, neither S2PL nor OCC would permit this execution, whereas SSI would allow it, because it contains only a single rw-antidependency.

SSI requires detecting rw-antidependencies at runtime. The SSI paper describes a method for identifying these dependencies by having transactions acquire locks in a special "SIREAD" mode on the data they read. These locks do not block conflicting writes (thus, "lock" is somewhat of a misnomer). Rather, a conflict between a SIREAD lock and a write lock flags an rw-antidependency, which might cause a transaction to be aborted. Furthermore, SIREAD locks must persist after a transaction commits, because conflicts can occur even after the reader has committed (*e.g.* the $T_1 \xrightarrow{rw} T_2$ conflict in Example 2). Corollary 2 implies that the locks must be retained until *all concurrent transactions* commit. Our PostgreSQL implementation uses SIREAD locks, but their implementation differs significantly because PostgreSQL was purely snapshot-based, as we describe in Section 5.2.

### 3.3.1 Variants on SSI

Subsequent work has suggested refinements to the basic SSI approach. Cahill's thesis [6] suggests a commit ordering optimization that can reduce false positives. Theorem 1 actually shows that every cycle contains a dangerous structure $T_1 \xrightarrow{rw} T_2 \xrightarrow{rw} T_3$, where *$T_3$ is the first to commit*. Thus, even if a dangerous structure is found, no aborts are necessary if either $T_1$ or $T_2$ commits before $T_3$. Verifying this condition requires tracking some additional state, but avoids some false positive aborts. We use an extension of this optimization in PostgreSQL. It does not, however, eliminate all false positives: there may not be a path $T_3 \rightsquigarrow T_1$ that closes the cycle. For example, in Example 2, if $T_1$'s REPORT accessed only the receipts table (not the current batch number), there would be no wr-dependency from $T_3$ to $T_1$, and the execution would be serializable with order $\langle T_1, T_2, T_3 \rangle$. However, the dangerous structure of rw-antidependencies $T_1 \xrightarrow{rw} T_2 \xrightarrow{rw} T_3$ would force some transaction to be spuriously aborted.

PSSI (Precisely Serializable Snapshot Isolation) is an extension of SSI that *does* eliminate all false positives [18]. It does so by building the full serialization history graph and testing it for cycles, rather than simply checking for dangerous structures. On a microbenchmark that stresses false aborts, PSSI can reduce the abort rate by up to 40% [18]. We considered this approach for PostgreSQL, but rejected it because we felt the costs outweighed the benefits of the reduced false positive abort rate. PSSI requires tracking wr- and ww-dependencies in addition to rw-antidependencies, consuming additional memory. Keeping the memory footprint small was an important requirement, and some of the optimizations we applied toward that end (Section 6) would not be compatible with PSSI. At the same time, the workloads we evaluate in Section 8 have a serialization failure rate well under 1%, suggesting additional precision has a limited benefit.

## 4. READ-ONLY OPTIMIZATIONS

Our version of SSI in PostgreSQL 9.1 includes new optimizations for read-only transactions. It's worthwhile to optimize specifically for read-only transactions: many workloads contain a significant fraction of read-only queries. Furthermore, long-running read-only transactions are also common. As we will discuss, these long-running transactions can substantially increase the overhead of SSI.

We improve performance for read-only transactions in two ways. Both derive from a new serializability theory result that characterizes when read-only transactions can be involved in SI anomalies. First, the theory enables a *read-only snapshot ordering optimization* to reduce the false-positive abort rate, an improved version of the commit ordering optimization described in Section 3.3.1. Second, we also identify certain *safe snapshots* on which read-only transactions can execute safely without any SSI overhead or abort risk, and introduce *deferrable transactions*, which delay their execution to ensure they run on safe snapshots.

### 4.1 Theory

Our read-only optimizations are based on the following extension of Theorem 1:

**Theorem 3.** *Every serialization anomaly contains a dangerous structure $T_1 \xrightarrow{rw} T_2 \xrightarrow{rw} T_3$, where if $T_1$ is read-only, $T_3$ must have committed before $T_1$ took its snapshot.*

*Proof.* Consider a cycle in the serialization history graph. From Theorem 1, we know it must have a dangerous structure $T_1 \xrightarrow{rw} T_2 \xrightarrow{rw} T_3$ where $T_3$ is the first transaction in the cycle to commit. Consider the case where $T_1$ is read-only.

Because there is a cycle, there must be some transaction $T_0$ that precedes $T_1$ in the cycle. (If the cycle has length 3, $T_0$ is the same transaction as $T_3$, but this does not affect the proof.) The edge $T_0 \rightarrow T_1$ can't be a rw-antidependency or a ww-dependency, because $T_1$ was read-only, so it must be a wr-dependency. A wr-dependency means that $T_0$'s changes were visible to $T_1$, so $T_0$ must have committed before $T_1$ took its snapshot. Because $T_3$ is the first transaction in the cycle to commit, it must commit before $T_0$ commits – and therefore before $T_1$ takes its snapshot. □

This result can be applied directly to reduce the false positive rate, using the following *read-only snapshot ordering rule*: if a dangerous structure is detected where $T_1$ is read-only, it can be disregarded as a false positive unless $T_3$ committed before $T_1$'s snapshot. Here, a transaction is considered read-only if it is explicitly declared as such (with BEGIN TRANSACTION READ ONLY) or if it has committed without modifying any data.

This result means that whether a read-only transaction can be a part of a dangerous structure depends only on when it takes its snapshot, not its commit time. Intuitively, it matters when read/write transactions commit, as this is the point when its changes become visible to other transactions. But it does not matter when read-only transactions commit, because they do not make any changes; only their snapshot times have an effect.

### 4.2 Safe Snapshots

If we can prove that a particular transaction will never be involved in a serialization anomaly, then that transaction can be run using standard snapshot isolation, without the need to track readsets for SSI. The rule above gives us a way to do so. A read-only transaction $T_1$ cannot have a rw-conflict pointing in, as it did not perform any writes. The only way it can be part of a dangerous structure, therefore, is if it has a conflict out to a concurrent read/write transaction $T_2$, and $T_2$ has a conflict out to a third transaction $T_3$ that committed before $T_1$'s snapshot. If no such $T_2$ exists, then $T_1$ will never cause a serialization failure. This depends only on the concurrent transactions, not on $T_1$'s behavior; therefore, we describe it as a property of the snapshot:

- **Safe snapshots:** A read-only transaction $T$ has a *safe snapshot* if no concurrent read/write transaction has committed with a rw-antidependency out to a transaction that committed before $T$'s snapshot, or has the possibility to do so.

A read-only transaction running on a safe snapshot can read any data (perform any query) without risk of serialization failure. It cannot be aborted, and does not need to take SIREAD locks. Conceptually, the set of transactions visible to a safe snapshot is a prefix of the apparent serial order of execution. This prevents precisely the situation in Figure 2. There, $T_1$ (the REPORT transaction) does not have a safe snapshot, because a concurrent transaction $T_2$ (NEW-RECEIPT) has a conflict out to an earlier transaction $T_3$ (CLOSE-BATCH). The conflict means $T_2$ must precede $T_3$ in the serial ordering. Because only $T_3$ is visible in $T_1$'s snapshot, its reads may (as in the example) contradict that serial ordering, requiring an abort.

An unusual property of this definition is that we cannot determine whether a snapshot is safe at the time it is taken, only once all concurrent read/write transactions complete, as those transactions might subsequently develop conflicts. Therefore, when a READ ONLY transaction is started, PostgreSQL makes a list of concurrent transactions. The read-only transaction executes as normal, maintaining SIREAD locks and other SSI state, until those transactions commit. After they have committed, if the snapshot is deemed safe, the read-only transaction can drop its SIREAD locks, essentially becoming a REPEATABLE READ (snapshot isolation) transaction. An important special case is a snapshot taken when no read/write transactions are active; such a snapshot is immediately safe and a read-only transaction using it incurs no SSI overhead.

### 4.3 Deferrable Transactions

Some workloads contain long-running read-only transactions. For example, one might run occasional analytic queries on a database that normally handles OLTP transactions. Periodic database maintenance tasks, such as backups using PostgreSQL's pg_dump utility, may also use long-running transactions. Such transactions are doubly problematic for SSI. Because they access large amounts of data, they take more SIREAD locks and are more likely to conflict with concurrent transactions. Worse, they inhibit cleanup of other transactions' SIREAD locks, because these locks must be kept until all concurrent transactions complete; this can easily exhaust memory.

These transactions would especially benefit from running on safe snapshots: they could avoid taking SIREAD locks, they would be guaranteed not to abort, and they would not prevent concurrent transactions from releasing their locks. *Deferrable transactions,* a new feature, provide a way to ensure that complex read-only transactions will always run on a safe snapshot. Read-only serializable transactions can be marked as deferrable with a new keyword, *e.g.* BEGIN TRANSACTION READ ONLY, DEFERRABLE. Deferrable transactions always run on a safe snapshot, but may block before their first query.

When a deferrable transaction begins, our system acquires a snapshot, but blocks the transaction from executing. It must wait for concurrent read/write transactions to finish. If any commit with a rw-conflict out to a transaction that committed before the snapshot, the snapshot is deemed unsafe, and we retry with a new snapshot. If all read/write transactions commit without such a conflict, the snapshot is deemed safe, and the deferrable transaction can proceed.

Note that deferrable transactions are not guaranteed to successfully obtain a safe snapshot within a fixed time. Indeed, for certain transaction patterns, it is possible that no safe snapshot ever becomes available. In theory, we could prevent this starvation by aborting concurrent transactions that would make the snapshot unsafe, or by preventing new transactions from starting. However, we have not found starvation to be a problem in practice. For example, in Section 8.4 we show that, even running concurrently with a heavy benchmark workload, deferrable transactions can usually obtain a safe snapshot within 1–6 seconds (and never more than 20 seconds).

## 5. IMPLEMENTING SSI IN POSTGRESQL

Our implementation of SSI – the first in a production database release – has some notable differences from previous implementations (as described in previous papers [6, 7, 18] and in Section 3.3). Much of these differences stem from the fact that PostgreSQL did not previously provide a true serializable isolation level. Previous implementations of SSI were built atop Berkeley DB [7] or MySQL's InnoDB [6, 18], both of which already supported strict two-phase locking. Accordingly, they were able to take advantage of features that were already present (*e.g.* predicate locking), whereas we needed to implement them anew. In particular, we had to build a new SSI lock manager; because it is designed specifically for tracking SIREAD locks, it has some unusual properties.

Our experience is especially relevant because SSI seems like a natural fit for databases like PostgreSQL that provide only snapshot-based isolation levels and lack a pre-existing serializable mode. One might expect SSI, being based on snapshot isolation, to be easier to implement on such databases than a traditional S2PL serializable level. We are the first to evaluate it in this context. As we discuss below, our experience suggests that SSI is actually more difficult to implement on such a database because it requires building much of the same lock manager infrastructure required to support S2PL.

### 5.1 PostgreSQL Background

Before delving into our SSI implementation, we begin by reviewing PostgreSQL's existing concurrency control mechanisms.

PostgreSQL previously provided two isolation levels – now three with the addition of SSI. Both were based on multiversion concurrency. The previous "SERIALIZABLE" mode provided snapshot isolation: every command in a transaction sees the same snapshot of the database, and write locks prevent concurrent updates to the same tuple. The weaker READ COMMITTED level essentially works the same way, but takes a new snapshot before each query rather than using the same one for the duration of the transaction, and handles concurrent updates differently. In PostgreSQL 9.1, the SERIALIZABLE level now uses SSI, and the snapshot isolation level remains available as REPEATABLE READ.

All queries in PostgreSQL are performed with respect to a snapshot, which is represented as the set of transactions whose effects are visible in the snapshot. Each tuple is tagged with the transaction ID of the transaction that created it (*xmin*), and, if it has been deleted or replaced with a new version, the transaction that did so (*xmax*). Checking which of these transactions are included in a snapshot determines whether the tuple should be visible. Updating a tuple is, in most respects, identical to deleting the existing version and creating a new tuple. The new tuple has a separate location in the heap, and may have separate index entries.[2] Here, PostgreSQL differs from other MVCC implementations (*e.g.* Oracle's) that update tuples in-place and keep a separate rollback log.

Internally, PostgreSQL uses three distinct lock mechanisms:

- **lightweight locks** are standard reader-writer locks for synchronizing access to shared memory structures and buffer cache pages; these are typically referred to as latches elsewhere in the literature
- **heavyweight locks** are used for long-duration (*e.g.* transaction-scope) locks, and support deadlock detection. A variety of lock modes are available, but normal-case operations such as SELECT and UPDATE acquire locks in non-conflicting modes. Their main purpose is to prevent schema-changing operations, such as DROP TABLE or REINDEX, from being run concurrently with other operations on the same table. These locks can also be explicitly acquired using LOCK TABLE.

---

[2]As an optimization, if an update does not modify any indexed fields, and certain other conditions hold, PostgreSQL may use a single index entry that points to a chain of tuple versions.

1855

- **tuple locks** prevent concurrent modifications to the same tuple. Because a transaction might acquire many such locks, they are not stored in the heavyweight lock table; instead, they are stored in the tuple header itself, reusing the *xmax* field to identify the lock holder. SELECT FOR UPDATE also acquires these locks. Conflicts are resolved by calling the heavyweight lock manager, to take advantage of its deadlock detection.

## 5.2 Detecting Conflicts

One of the main requirements of SSI is to be able to detect rw-conflicts as they happen. Earlier work suggested modifying the lock manager to acquire read locks in a new SIREAD mode, and flagging a rw-antidependency when a conflicting lock is acquired. Unfortunately, this technique cannot be directly applied to PostgreSQL because the lock managers described above do not have the necessary information. To begin with, PostgreSQL did not previously acquire read locks on data accessed in *any* isolation level, unlike the databases used in prior SSI implementations, so SIREAD locks cannot simply be acquired by repurposing existing hooks for read locks. Worse, even with these locks, there is no easy way to match them to conflicting write locks because PostgreSQL's tuple-level write locks are stored in tuple headers on disk, rather than an in-memory table.

Instead, PostgreSQL's SSI implementation uses existing MVCC data as well as a new lock manager to detect conflicts. Which one is needed depends on whether the write happens chronologically before the read, or vice versa. If the write happens first, then the conflict can be inferred from the MVCC data, without using locks. Whenever a transaction reads a tuple, it performs a visibility check, inspecting the tuple's *xmin* and *xmax* to determine whether the tuple is visible in the transaction's snapshot. If the tuple is not visible because the transaction that created it had not committed when the reader took its snapshot, that indicates a rw-conflict: the reader must appear before the writer in the serial order. Similarly, if the tuple has been deleted – *i.e.* it has an *xmax* – but is still visible to the reader because the deleting transaction had not committed when the reader took its snapshot, that is also a rw-conflict that places the reader before the deleting transaction in the serial order.

We also need to handle the case where the read happens before the write. This cannot be done using MVCC data alone; it requires tracking read dependencies using SIREAD locks. Moreover, the SIREAD locks must support predicate reads. As discussed earlier, none of PostgreSQL's existing lock mechanisms were suitable for this task, so we developed a new SSI lock manager. The SSI lock manager stores only SIREAD locks. It does not support any other lock modes, and hence cannot block. The two main operations it supports are to obtain a SIREAD lock on a relation, page, or tuple, and to check for conflicting SIREAD locks when writing a tuple.

### 5.2.1 Implementation of the SSI Lock Manager

The PostgreSQL SSI lock manager, like most lock managers used for S2PL-based serializability, handles predicate reads using index-range locks (in contrast to actual predicate locks [9]). Reads acquire SIREAD locks on all tuples they access, and index access methods acquire SIREAD locks on the "gaps" to detect phantoms. Currently, locks on B+-tree indexes are acquired at page granularity; we intend to refine this to next-key locking [16] in a future release. Both heap and index locks can be promoted to coarser granularities to save space in the lock table, *e.g.* replacing multiple tuple locks with a single page lock.

One simplification we were able to make is that intention locks were not necessary, despite the use of multigranularity locking (and contrary to a suggestion that intention-SIREAD locks would be required [7]). It suffices to check for locks at each granularity (relation, page, and tuple) when writing a tuple. To prevent problems with concurrent granularity promotion, these checks must be done in the proper order: coarsest to finest.

Some other simplifications arise because SIREAD locks cannot cause blocking. Deadlock detection becomes unnecessary, though this was not a significant benefit because PostgreSQL already had a deadlock detector. It also simplifies placement of the calls to acquire locks and check for conflicts. In a traditional lock implementation, these calls must be carefully placed where no lightweight locks are held (*e.g.* no buffer pool pages are locked), because blocking while these are held might cause a lock-latch deadlock.

However, the SSI lock manager must also handle some situations that a typical S2PL lock manager does not. In particular, SIREAD locks must be kept up to date when concurrent transactions modify the schema with data-definition language (DDL) statements. Statements that rewrite a table, such as RECLUSTER or ALTER TABLE, cause the physical location of tuples to change. As a result, page- or tuple-granularity SIREAD locks, which are identified by physical location, are no longer valid; PostgreSQL therefore promotes them to relation-granularity. Similarly, if an index is removed, any index-gap locks on it can no longer be used to detect conflicts with a predicate read, so they are replaced with a relation-level lock on the associated heap relation. These issues don't arise in a S2PL lock manager, as holding a read lock on a tuple would block the DDL operations described here until the reading transaction completes. SIREAD locks, however, are retained after a transaction commits, so it would be overly restrictive if they blocked DDL operations.

## 5.3 Tracking Conflicts

The previous section described how to detect rw-antidependencies, but one antidependency alone is not a problem; it is only a dangerous structure of two rw-antidependencies that may cause an anomaly. Detecting when this is the case requires keeping some state to represent serializable transactions and their dependencies.

One question we were faced with was how much information to track about a transaction's dependencies. Each previous SSI implementation has answered this question differently. The original SSI paper suggested two single-bit flags per transaction: whether the transaction had a rw-antidependency pointing in, and whether it had one pointing out [7]. Later, this was extended to two pointers, with a pointer-to-self being used to represent a transaction with multiple rw-antidependencies in or out [6]. PSSI opted instead to store the entire graph, including wr- and ww-dependencies, to support cycle-testing [18].

We chose to keep a list of all rw-antidependencies in or out for each transaction, but not wr- and ww-dependencies. Keeping pointers to the other transaction involved in the rw-antidependency, rather than a simple flag, is necessary to implement the commit ordering optimization described in Section 3.3 and the read-only optimization of Section 4.1. It also allows us to remove conflicts if one of the transactions involved has been aborted. Keeping only one pointer would require us to abandon these optimization for transactions with multiple rw-antidependencies in or out. We also implemented a number of techniques to aggressively discard information about committed transactions to conserve memory (Section 6), and these require accurate information about the rw-antidependency graph.

We considered the PSSI approach, which uses cycle testing to eliminate false positives, but did not use it because it requires tracking ww- and wr-dependencies. As mentioned above, we were concerned about memory usage, so we did not want to track additional dependencies. More fundamentally, we were concerned about wr-dependencies that take place partially outside the database, which we cannot track. For example, an alternate implementation of the batch processing example might implement the REPORT operation described in Section 2 as two separate transactions: one that queries the batch number and another that obtains all receipts for a particular batch. A user might run one transaction that reads the batch number and observes that batch *x* is current, and then – in a separate transaction – list the receipts for batch *x* − 1. Having observed the effects of the CLOSE-BATCH transaction that incremented the batch number, the user could reasonably expect that no further receipts would be added for the closed batch. PSSI, however, would not detect this dependency (as it was a separate transaction that read the batch number) and allow an anomaly similar to the one in Figure 2. This problem could be resolved by tracking causal dependencies between transactions: a transaction should not appear to execute before a previous transaction from the same user. However, properly tracking causal dependencies between multiple communicating clients requires substantial support from the application.

### 5.4 Resolving Conflicts: Safe Retry

When a dangerous structure is found, and the commit ordering conditions are satisfied, some transaction must be aborted to prevent a possible serializability violation. It suffices to abort any one of the transactions involved (unless it has already committed). We want to choose the transaction to abort in a way that ensures the following property:

- **Safe retry:** if a transaction is aborted, immediately retrying the same transaction will not cause it to fail again with the same serialization failure.

The safe retry property is desirable because it prevents wasted work from repeatedly retrying the same transaction, particularly in a configuration we expect to be common: using a middleware layer to automatically retry transactions aborted for serialization failures.

Once we have identified a dangerous structure $T_1 \xrightarrow{rw} T_2 \xrightarrow{rw} T_3$, the key principle for ensuring safe retry is to abort a transaction that conflicts with a *committed* transaction. When the aborted transaction is retried, it will not be concurrent with the committed transaction, and cannot conflict with it. Specifically, the following rules are used to ensure safe retry:

1. Do not abort anything until $T_3$ commits. This rule is needed to support the commit ordering optimization, but it also serves the safe retry goal.
2. Always choose to abort $T_2$ if possible, *i.e.* if it has not already committed. $T_2$ must have been concurrent with both $T_1$ and $T_3$. Because $T_3$ is already committed, the retried $T_2$ will not be concurrent with it and so will not be able to have a rw-conflict out to it, preventing the same error from recurring. (If we had chosen to abort $T_1$ instead, it would still be concurrent with $T_2$, so the same dangerous structure could form again.)
3. If both $T_2$ and $T_3$ have committed when the dangerous structure is detected, then the only option is to abort $T_1$. But this is safe; $T_2$ and $T_3$ have already committed, so the retried transaction will not be concurrent with them, and cannot conflict with either.

Note that rule (1) means that dangerous structures may not be resolved immediately when they are detected. As a result, we also perform a check when a transaction commits. If $T_3$ attempts to commit while part of a dangerous structure of uncommitted transactions, it is the first to commit and an abort is necessary. This should be resolved by aborting $T_2$, for the same reasoning as in (2).

One might worry that this delayed resolution could cause wasted work or additional conflicts, because a transaction continues to execute even after a conflict that could force it to abort. However, aborting a transaction immediately would cause an equivalent amount of wasted work, if the transaction is immediately retried only to abort again. In fact, the delayed resolution is *less* wasteful because it may ultimately not be necessary to abort transactions at all, depending on the order in which they commit.

These rules become slightly more complex when two-phase commit is involved, and safe retry may be impossible, an issue we discuss in Section 7.1.

## 6. MEMORY USAGE MITIGATION

After implementing the basic SSI functionality, one of the problems we were immediately confronted with was its potentially unbounded memory usage. The problem is not merely that one transaction can hold a large number of locks – a standard lock manager problem – but one unique to SSI: a transaction's locks cannot be released until that transaction *and* all concurrent transactions commit. Moreover, other transaction state (the rw-antidependency graph) may need to be retained even longer to check for dangerous structures. Thus, a single long-running transaction can easily prevent thousands of transactions from being cleaned up.

We were faced with two requirements related to memory usage. The SSI implementation's memory usage must be *bounded*: the lock table and dependency graph must have a fixed size (specified by the configuration file). The system must also be able to *gracefully degrade*. Even in the presence of long-running transactions, the system should not fail to process new transactions because it runs out of memory. Instead, it should be able to accept new transactions, albeit possibly with a higher false positive abort rate.

These requirements were driven in part by PostgreSQL's restrictive limitations on shared memory. PostgreSQL stores all its shared memory in a single System V shared memory segment. The default configuration of many operating systems restricts the size of this segment (*e.g.* to 32 MB on Linux), so SSI must be able to function even in a low-memory scenario. PostgreSQL also lacks effective support for dynamic allocation of shared memory, forcing us to allocate a fixed amount of memory for the lock table at startup. However, the problem is not PostgreSQL-specific; although other databases might be less likely to exhaust shared memory, any memory used for storing SSI state is memory that cannot put to more productive uses, such as the buffer cache.

Our PostgreSQL implementation uses four techniques to limit the memory usage of the SSI lock manager. We have already seen the first two; the others are discussed below:

1. Safe snapshots and deferrable transactions (Section 4.2) can reduce the impact of long-running read-only transactions
2. Granularity promotion (Section 5.2): multiple fine-grained locks can be combined into a single coarse-grained lock to reduce space.
3. *Aggressive cleanup* of committed transactions: the parts of a transaction's state that are no longer needed after commit are removed immediately
4. *Summarization* of committed transactions: if necessary, the state of multiple committed transactions can be consolidated into a more compact representation, at the cost of an increased false positive rate

### 6.1 Aggressive Cleanup

How long does information about a committed transaction need to be retained? As mentioned previously, a committed transaction's SIREAD locks are no longer necessary once all concurrent trans-

1857

actions have committed, as only concurrent transactions can be involved in a rw-antidependency. Therefore, we clean up unnecessary locks when the oldest active transaction commits. However, some information about the conflict graph must be retained longer. Specifically, if an active transaction $T_1$ develops a conflict out to a committed transaction $T_2$, we need to know whether $T_2$ has a conflict out to a third transaction $T_3$, and $T_3$'s commit sequence number. But $T_3$ may have committed before any active transaction began, meaning that it was already cleaned up. To prevent this problem, we record an additional piece of information in each transaction's node: the commit sequence number of the earliest committed transaction to which it has a conflict out.

We use another optimization when the only remaining active transactions are read-only. In this case, the SIREAD locks of all committed transactions can be safely discarded. Recall that SIREAD locks are only needed to detect conflicts when a concurrent transaction's write happens after another transaction's read – and there are no active transactions that can write. Furthermore, the committed transactions' lists of rw-antidependencies *in* can be discarded, because these dependencies could only be a part of a dangerous structure if an active read-write transaction modified some object read by the committed transaction.

## 6.2 Summarizing Committed Transactions

Our SSI implementation reserves storage for a fixed number of committed transactions. If more committed transactions need to be tracked, we *summarize* the state of previously committed transactions. It is usually sufficient to discover that a transaction has a conflict with *some* previously committed transaction, but not which one. Summarization allows the database to continue accepting new transactions, although the false positive abort rate may increase because some information is lost in the process.

Our summarization procedure is based on the observation that information about committed transactions is needed in two cases:

First, an active transaction modifying a tuple needs to know if some committed transaction read that tuple. This could create a dangerous structure $T_{committed} \xrightarrow{rw} T_{active} \xrightarrow{rw} T_3$. We need to keep a SIREAD lock to detect that such a transaction existed – but it does not matter what specific transaction it was, whether it had other rw-antidependencies in or out, etc. This motivates the first part of summarizing a committed transaction: the summarized transaction's SIREAD locks are *consolidated* with those of other summarized transactions, by reassigning them to a single dummy transaction. Each lock assigned to this dummy transaction also records the commit sequence number of the most recent transaction that held the lock, to determine when the lock can be cleaned up. The benefit of consolidation is that each lock only needs to be recorded once, even if it was held by multiple committed transactions. Combined with the ability to promote locks to a coarser granularity, this can make it unlikely that the SIREAD lock table will be exhausted.

Second, an active transaction reading a tuple needs to know whether that tuple was written by a concurrent serializable transaction. This could create one of two possible dangerous structures:

$$T_1 \xrightarrow{rw} T_{active} \xrightarrow{rw} T_{committed} \quad \text{or} \quad T_{active} \xrightarrow{rw} T_{committed} \xrightarrow{rw} T_3$$

Recall that we detect this situation using the transaction ID of the writer that is stored in the tuple header (Section 5.2). However, we still need to check whether that the writer was a serializable transaction, as opposed to one running with a weaker isolation level. Furthermore, we need to know whether that transaction had a conflict out to a third transaction $T_3$ (and whether $T_3$ committed first), to detect the second case above. For non-summarized transactions, this information is available from the dependency graph. Summarized transactions, however, are removed from the graph. Instead, we keep a simple table mapping a summarized transaction's ID to the commit sequence number of the oldest transaction to which it has a conflict out. This can be represented using a single 64 bit integer per transaction, and the table can be swapped out to disk using an existing LRU mechanism in PostgreSQL, giving it effectively unlimited capacity.

## 7. FEATURE INTERACTIONS

PostgreSQL has a wide variety of features, some of which have interesting or unexpected interactions with SSI. We describe several such interactions in this section. To our knowledge, previous implementations of SSI have not addressed these issues.

### 7.1 Two-Phase Commit

PostgreSQL supports two-phase commit: the PREPARE TRANSACTION command ensures a transaction is stored on disk, but does not make its effects visible.[3] A subsequent COMMIT PREPARED is guaranteed to succeed, even if the database server crashes and recovers in the meantime. This requires writing the list of locks held by the transaction to disk, so that they will persist after recovery.

We extended this procedure so that a transaction's SIREAD locks will also be written to disk; they, too, must persist after a crash/recovery, because the transaction remains active after recovery and new concurrent transactions may conflict with it. PostgreSQL also needs to know, after recovery, whether the prepared transaction had any rw-antidependencies in or out. It isn't feasible, however, to record that information in a crash-safe way: the dependency graph could be large, and new conflicts may be detected even after the transaction prepares. Accordingly, after a crash, we conservatively assume that any prepared transaction has rw-antidependencies both in and out.

A transaction that has PREPARED cannot be aborted. This means that we must perform the pre-commit serialization failure check described in Section 5.4 before preparing. It also means that any serialization failures involving a prepared transaction must be resolved by aborting one of the other transactions involved. Unfortunately, this sometimes makes it impossible to guarantee the safe retry property of Section 5.4. Consider a dangerous structure involving an active transaction, a prepared transaction, and a committed one:

$$T_{active} \xrightarrow{rw} T_{prepared} \xrightarrow{rw} T_{committed}$$

According to the safe retry rules in Section 5.4, we should choose to abort the "pivot" transaction, $T_{prepared}$ – but we cannot, as it has prepared. Our only option is to abort $T_{active}$ instead. If the user immediately retries that transaction, however, it will still be concurrent with (and can still conflict with) $T_{prepared}$, as $T_{prepared}$ has not yet committed. The retried transaction is therefore likely to be aborted because of the same conflict.

### 7.2 Streaming Replication

Beginning with last year's 9.0 release, PostgreSQL has built-in support for master-slave replication. As in many other database systems, this is implemented using log shipping: the master streams write-ahead-log records to the slave, which can process read-only transactions while it applies the updates from the master.

Unfortunately, log-shipping replication does not provide serializable behavior when used with SSI and read-only transactions on the slaves. Two-phase locking has the property that the commit order of transactions matches the apparent serial order; the same is true of the standard optimistic concurrency control technique [15].

---

[3]By design, PostgreSQL does not itself support distributed transactions; its two-phase commit support is intended as a primitive that can be used to build an external transaction coordinator.

As a result, running read-only queries on a snapshot of the database guarantees serializability without locking. SSI does not have this property, however. With SSI, a read-only query can cause serialization failures: recall the batch-processing example's REPORT transaction in Section 2. If run on the master, SSI would detect the rw-antidependencies between that transaction and the others, and abort one of them. If, however, that REPORT transaction were run on a slave replica, the dependency would go unnoticed and the transaction would see anomalous results. A similar problem could occur with a snapshot-based transactional cache [17].

Currently, PostgreSQL does not allow serializable transactions to be run on the slaves. We plan to eliminate this restriction in a future release. We quickly discounted the option of attempting to track rw-antidependencies caused by queries on slave replicas. This would require the slaves to communicate information back to the master about their transactions' read sets. The cost and complexity of doing so, along with the required synchronization, would likely eliminate much of the benefit of running queries on the slave.

Instead, we use our notion of safe snapshots (Section 4.2), on which we *can* run any read-only query. Slave replicas will run serializable transactions only on safe snapshots, eliminating the need for them to track read dependencies or communicate them to the master. We plan to achieve this by adding information to the log stream that identifies safe snapshots. Then, transactions running on the slaves will have one of three options: they can use the most recent (but potentially stale) safe snapshot; they can wait for the next available safe snapshot (as DEFERRABLE transactions do); or they can simply run at a weaker isolation level (as is possible now).

### 7.3 Savepoints and Subtransactions

PostgreSQL uses subtransactions to implement savepoints, an issue not addressed by previous SSI implementations. Creating a savepoint starts a new nested subtransaction. This subtransaction cannot commit until the top-level transaction commits. However, it can be aborted using the ROLLBACK TO SAVEPOINT command, discarding all changes made since the savepoint. For the most part, subtransactions have little impact on SSI. We do not drop SIREAD locks acquired during a subtransaction if the subtransaction is aborted (*i.e.* all SIREAD locks belong to the top-level transaction). This is because data read during the subtransaction may have been reported to the user or otherwise externalized.

However, subtransactions interact poorly with an optimization not previously discussed. As suggested in Cahill's thesis [6], we allow a transaction to drop its SIREAD lock on a tuple if it later modifies that tuple. This optimization is safe because the write lock is held until the transaction commits, preventing concurrent transactions from modifying the same tuple and thereby obviating the need for the SIREAD lock. It is a particularly useful optimization in PostgreSQL because the write lock is stored directly in the tuple header, so storing it has effectively no cost beyond that of updating the tuple, whereas SIREAD locks must be stored in an in-RAM table. This optimization cannot be used while executing a subtransaction, because the write lock is associated with the subtransaction. If that subtransaction is rolled back, the write lock will be released, leaving the top-level transaction without either a write or SIREAD lock.

### 7.4 Index Types

Our discussion of predicate locking has focused mainly on B$^+$-trees, the most common index type. PostgreSQL provides several other types of built-in indexes, including GiST [13] and GIN indexes, and supports an extensible index API, allowing users to define their own index access methods [20]. In general, new index access methods must indicate whether they support predicate locking; if so, they are required to acquire the appropriate SIREAD locks to avoid phantoms. Otherwise, PostgreSQL falls back on acquiring a relation-level lock on the index whenever it is accessed.

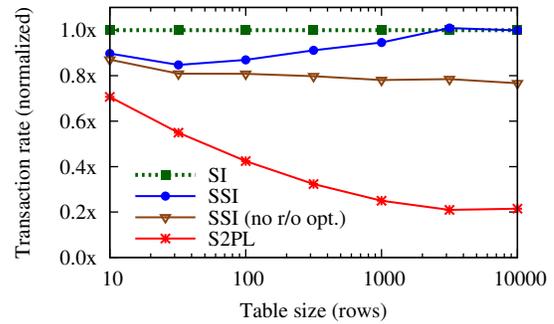

Figure 4: SIBENCH transaction throughput for SSI and S2PL as a percentage of SI throughput

Of PostgreSQL's built-in index access methods, currently only B$^+$-trees support predicate locking. We plan to add support for GiST indexes in an upcoming release, following a similar approach; the major difference is that GiST indexes must lock internal nodes in the tree, while B$^+$-tree indexes only lock leaf pages. Support for GIN and hash indexes is also planned.

## 8. EVALUATION

Running transactions in PostgreSQL's SERIALIZABLE mode comes with a performance cost, compared to snapshot isolation. There are two sources of overhead. First, tracking read dependencies and maintaining the serialization graph imposes CPU overhead and can cause contention on the lock manager's lightweight locks. Second, transactions may need to be retried after being aborted by serialization failures, some of which may be false positives.

In this section, we evaluate the cost of serializability in PostgreSQL 9.1. We compare the performance of our SSI implementation to PostgreSQL's existing snapshot isolation level (REPEATABLE READ). To provide additional context, we also compare with a simple implementation of strict two-phase locking for PostgreSQL. This implementation reuses our SSI lock manager's support for index-range and multigranularity locking; rather than acquiring SIREAD locks, it instead acquires "classic" read locks in the heavyweight lock manager, as well as the appropriate intention locks.

We evaluated the performance on PostgreSQL on three workloads: the SIBENCH microbenchmark (Section 8.1), a modified TPC-C-like transaction processing benchmark (Section 8.2), and the RUBiS web application benchmark (Section 8.3). We used several hardware configurations to test both CPU and disk bottlenecks. In each case, PostgreSQL's settings were tuned for the hardware using pgtune.[4]

### 8.1 SIBENCH Microbenchmark

SIBENCH is a simple microbenchmark that demonstrates the benefit of snapshot isolation and SSI over locking approaches when there are many rw-conflicts [6]. The database consists of a single table containing $N$ ⟨key, value⟩ pairs. The SIBENCH workload consists of equal numbers of update transactions, which update the value for one randomly-selected key, and query transactions, which scan the entire table to find the key with the lowest value. We ran this benchmark on a 2.83 GHz Core 2 Quad Q9550 system with 8 GB RAM running Ubuntu 11.10. The database was stored on an in-

---
[4] http://pgtune.projects.postgresql.org/

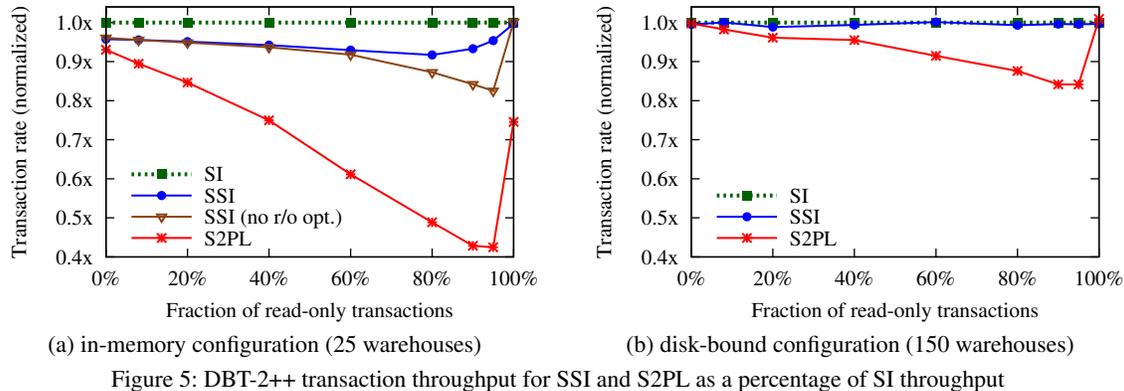

(a) in-memory configuration (25 warehouses)  (b) disk-bound configuration (150 warehouses)

Figure 5: DBT-2++ transaction throughput for SSI and S2PL as a percentage of SI throughput

memory file system (tmpfs), so that the benchmark can measure CPU overhead and contention caused by concurrency control.

Figure 4 shows the throughput in committed transactions per second for SSI and S2PL, relative to the performance of snapshot isolation. Locking imposes a clear performance penalty compared to SI, as update transactions cannot run concurrently with query transactions. SSI obtains throughput close to that of SI because it permits them to execute concurrently. On this simple benchmark, tracking read dependencies has a CPU overhead of 10–20%. Our read-only optimizations reduce this cost. A query transaction can be determined to have a safe snapshot once any update transactions that were active when it started complete; thereafter, it does not have to track read dependencies. For larger table sizes, the query transactions run longer, making this more likely.

## 8.2 Transaction Processing: DBT-2++

To measure the overhead of our implementation of SSI using a more realistic workload, we used DBT-2,[5] an open-source transaction processing benchmark inspired by TPC-C [21]. TPC-C is known not to exhibit anomalies under snapshot isolation [10], so we incorporated the "credit check" transaction from Cahill's "TPC-C++" variant, which can create a cycle of dependencies when run concurrently with other transactions [6]. We also applied some optimizations to eliminate frequent points of contention, including caching certain read-only data and omitting the warehouse year-to-date totals.

We ran DBT-2++ in two configurations to measure the overhead of SSI. First, we used a 25-warehouse scale factor (a 3 GB dataset) and used the in-memory tmpfs system described above; this allows us to measure the CPU overhead of tracking dependencies in a "worst-case" CPU-bound environment. The second configuration used a 150-warehouse (19 GB) database and was disk-bound, allowing us to evaluate the rate of serialization failures in a configuration with more concurrent and longer-running transactions. For this configuration, we used a 16-core 1.60 GHz Xeon E7310 system with 8 GB of RAM running Ubuntu 11.04. The database was stored on a 3-disk RAID 5 array of Fujitsu MAX3073RC 73 GB 15,000 RPM drives, with an identical fourth drive dedicated to PostgreSQL's write-ahead log. The RAID controller used a battery-backed write-back cache.

The standard TPC-C workload mix consists of 8% read-only transactions. To gain further insight, we scaled the workload mix to contain different fractions of read-only transactions, keeping the transaction proportions otherwise identical. We used concurrency levels of 4 and 36 threads on the in-memory and disk-bound workloads respectively, as these achieved the highest performance. We ran the benchmark with no think time, and measured the resulting throughput, shown in Figure 5. Again, the performance of SSI and S2PL is shown relative to the performance of snapshot isolation.

[5] http://osdldbt.sourceforge.net

|      | Throughput (req/s) | Serialization failures |
|------|--------------------|------------------------|
| SI   | 435                | 0.004%                 |
| SSI  | 422                | 0.03%                  |
| S2PL | 208                | 0.76%                  |

Figure 6: RUBiS performance

For the in-memory configuration (Figure 5a), SSI causes a 5% slowdown relative to snapshot isolation because of increased CPU usage. Our read-only optimizations reduce the CPU overhead of SSI for workloads with mostly read-only transactions. SSI outperforms S2PL for all transaction mixes, and does so by a significant margin when the fraction of read-only transactions is high. On these workloads, there are more rw-conflicts between concurrent transactions, so locking imposes a larger performance penalty. (The 100%-read-only workload is a special case; there are no lock conflicts under S2PL, and SSI has no overhead because all snapshots are safe.) The 150-warehouse configuration (Figure 5b) behaves similarly, but the differences are less pronounced: on this disk-bound benchmark, CPU overhead is not a factor, and improved concurrency has a limited benefit. Here, the performance of SSI is indistinguishable from that of SI. Transactions rarely need to be retried; in all cases, the serialization failure rate was under 0.25%.

## 8.3 Application Performance: RUBiS

We also measured the impact of SSI on application-level performance using the RUBiS web application benchmark [3]. RUBiS simulates an auction site modeled on eBay. We used the PHP implementation of RUBiS, configured with the standard "bidding" workload (85% read-only and 15% read/write transactions), and a dataset containing 225,000 active auctions, 1 million completed auctions, and 1.35 million users, for a total database size of 6 GB.

In these benchmarks, the database server ran on a 2.83 GHz Core 2 Quad Q9550 system with 8 GB RAM and a Seagate ST3500418AS 500 GB 7200 RPM hard drive running Ubuntu 11.10. Application server load can be a bottleneck on this workload [3], so we used multiple application servers (running Apache 2.2.17 and PHP 5.3.5) so that database performance was always the limiting factor.

The RUBiS workload contains frequent rw-conflicts. For example, queries that list the current bids on all items in a particular category conflict with requests to bid on those items. Accordingly, two-phase locking incurs significant overhead from lock contention, as seen in Figure 6. Furthermore, deadlocks occasionally occur, requiring expensive deadlock detection and causing serialization failures. SSI achieves performance comparable to snapshot isolation, because dangerous structures are rare and so transactions are rarely aborted.

## 8.4 Deferrable Transactions

In Section 4.3, we introduced deferrable transactions. Aimed at long-running analytic queries, this feature allows transactions to avoid the overhead of SSI by running them under snapshot isolation on a safe snapshot. The tradeoff is that these transactions may have to wait until a safe snapshot is detected.

How long it takes to obtain a safe snapshot depends on what transactions are running concurrently. We tested deferrable transactions with the DBT-2++ workload described above (using the disk-bound configuration and the standard 8% read-only transactions). This produces a heavy load with many concurrent transactions, making it a particularly challenging case for deferrable transactions. While the benchmark was executing, we started a deferrable transaction, ran a trivial query, and measured how long it took to find a safe snapshot. We repeated this 1200 times with a one-second delay between deferrable transactions. The median latency was 1.98 seconds, with 90% of transactions able to obtain a safe snapshot within 6 seconds, and all within 20 seconds. Given the intended use (long-running transactions), we believe this delay is reasonable.

## 9. CONCLUSION

Serializable transactions can simplify development by allowing database users to ignore concurrency issues – an advantage that becomes particularly relevant with today's highly-concurrent systems. Despite this, many users are unwilling to pay any performance cost for serializability. For this reason, PostgreSQL historically did not even provide serializability, instead offering snapshot isolation as its highest isolation level. We addressed this in PostgreSQL 9.1 with a new, SSI-based serializable isolation level. Our experiments show that this serializable mode provides performance similar to snapshot isolation and considerably outperforms strict two-phase locking on read-intensive workloads – hopefully making it a practical option for developers who would have previously opted to use snapshot isolation and endure the resulting anomalies.

Our implementation of SSI is the first in production use, as well as the first in a database that did not previously provide a serializable isolation level. This presented us with a number of new challenges. We had to implement a new predicate lock manager that tracks the read dependencies of transactions, integrate SSI with existing PostgreSQL features, and develop a transaction summarization technique to bound memory usage. We also introduced new optimizations for read-only transactions. Given these challenges, implementing SSI proved more challenging than a typical S2PL-based serializable mode. Despite this, the resulting performance benefits made the effort worthwhile, and played a key role in making this serializable isolation level acceptable to the PostgreSQL community.

## 10. ACKNOWLEDGMENTS

Developing the SSI-based serializable isolation level and integrating it into the PostgreSQL release required support from many members of the PostgreSQL community. Heikki Linnakangas, Robert Haas, Jeff Davis, and Joe Conway reviewed the code and provided valuable advice. Anssi Kääriäinen and Yamamoto Takahashi found numerous bugs during careful testing. Markus Wanner and Heikki Linnakangas implemented testing frameworks that proved invaluable during development. We also thank James Cowling, Evan Jones, Barbara Liskov, Sam Madden, David Schultz, and Irene Zhang for helpful feedback about this paper. Dan Ports was supported in part by the National Science Foundation under grant CNS-0834239.